\documentclass{emulateapj}

\usepackage{amssymb}
\usepackage{amsmath}
\usepackage{graphicx}
\usepackage{rotating}

\newcommand{\subscript}[1]{\textnormal{\tiny{#1}}}
\newcommand{\tk}{\ensuremath{T_\subscript{K}}}
\newcommand{\tvib}{\ensuremath{T_\subscript{vib}}}

\newcommand{\td}{\ensuremath{T_\subscript{dust}}}
\newcommand{\rstar}{\ensuremath{R_\star}}
\newcommand{\kms}{km~s$^{-1}$}
\newcommand{\cm}{cm$^{-1}$}
\newcommand{\cmmm}{cm$^{-3}$}
\newcommand{\mlr}{M$_\odot$~yr$^{-1}$}
\newcommand{\vexp}{\ensuremath{v_\subscript{exp}}}

\newcommand{\irc}{IRC+10216}

\newcommand{\sci}{Sci}
\newcommand{\jai}{JAI}
\newcommand{\aaprev}{Astron. Astrophys. Rev}

\begin{document}

\title{Detection of the S(1) Rotational Line of H$_2$ toward \irc: A Simultaneous Measurement of Mass-Loss Rate and CO Abundance}
\shorttitle{The pure rotational line S(1) of H$_2$ toward \irc}
\shortauthors{J. P. Fonfr\'{\i}a et al.}
\author{J.~P.~Fonfr\'ia\altaffilmark{1}}
\author{C.~N.~DeWitt\altaffilmark{2}}
\author{E.~J.~Montiel\altaffilmark{2}}
\author{J.~Cernicharo\altaffilmark{1}}
\author{M.~J.~Richter\altaffilmark{3}}
\affiliation{
  $^1$Molecular Astrophysics Group, Instituto de F\'isica Fundamental, IFF-CSIC, C/ Serrano, 123, 28006, Madrid (Spain)\\
  $^2$ SOFIA-USRA, NASA Ames Research Center, MS 232-12, Moffett Field, CA 94035 (USA) \\
  $^3$Physics Dept. - UC Davis, One Shields Ave., Davis, CA 95616 (USA)
}

\email{jpablo.fonfria@csic.es}

\begin{abstract}
  We report the first detection of the S(1) pure rotational line of ortho-H$_2$ at 17.04~$\mu$m in an asymptotic giant branch star, using observations of \irc{} with the Echelon-cross-echelle Spectrograph (EXES) mounted on the Stratospheric Observatory for Infrared Astronomy (SOFIA).
  This line, which was observed in a very high sensitivity spectrum (RMS noise~$\simeq 0.04\%$ of the continuum), was detected in the wing of a strong telluric line and displayed a P Cygni profile.
  The spectral ranges around the frequencies of the S(5) and S(7) ortho-H$_2$ transitions were observed as well but no feature was detected in spectra with sensitivities of 0.12\% and 0.09\% regarding the continuum emission, respectively.
  We used a radiation transfer code to model these three lines and derived a mass-loss rate of $(2.43\pm 0.21)\times 10^{-5}$~\mlr{} without using the CO abundance.
  The comparison of this rate with previous estimates derived from CO observations suggests that the CO abundance relative to H$_2$ is $(6.7\pm 1.4)\times 10^{-4}$.
From this quantity and previously reported molecular abundances, we estimate the O/H and C/H ratios to be $(3.3\pm 0.7)\times 10^{-4}$ and $>(5.2\pm 0.9)\times 10^{-4}$, respectively.
The C/O ratio is $>1.5\pm 0.4$.
  The absence of the S(5) and S(7) lines of ortho-H$_2$ in our observations can be explained by the opacity of hot dust within 5\rstar{} from the center of the star.
  We estimate the intensity of the S(0) and S(2) lines of para-H$_2$ to be $\simeq 0.1\%$ and 0.2\% of the continuum, respectively, which are below the detection limit of EXES.
\end{abstract}

\maketitle

\section{Introduction}
\label{sec:introduction}

Asymptotic giant branch (AGB) and red supergiant stars eject large amounts of processed matter to the interstellar medium by means of winds composed of dust grains and molecular gas.
These winds build the circumstellar envelopes (CSEs), where different structures, such as spirals, arcs, and detached shells related to episodic mass-loss and thermal pulses are frequently spotted \citep{mauron_2006,maercker_2014,guelin_2018,decin_2020}.
Many molecules have been detected toward these CSEs, mostly owing to the photo-dissociation in their outer layers of molecular species formed closer to the star \citep{cernicharo_2000,zack_2011,anderson_2014,agundez_2017,velilla-prieto_2017}.
In all cases, the mass-loss rate is a crucial parameter to understand the kinematic and chemical evolution of the CSEs.
Moreover, it is thought to impact the stellar evolution if it exceeds the nucleosynthesis rate \citep{hofner_2018}.

The most abundant molecule in the inner layers of CSEs is H$_2$ followed by CO, H$_2$O, C$_2$H$_2$, N$_2$, SiO, HCN, and other compounds with abundances relative to H$_2$ typically below $10^{-5}$ \citep[e.g.,][]{agundez_2020}.
In spite of the high kinematic and thermodynamic importance of dust grains, they account for a negligible amount of matter related to the total \citep[$\sim$0.1-1.0\%;][]{schoier_2001,groenewegen_2002,ramstedt_2009,danilovich_2015}.
Thus, the best way to determine the mass-loss rate of a star would be to observe the H$_2$ spectrum.
However, electric dipole transitions are forbidden for this molecule and its electric quadrupole lines are very weak.

Ro-vibrational lines of H$_2$ at 2~$\mu$m were detected toward cool stars long ago \citep[e.g.,][]{johnson_1983,lambert_1986,hinkle_2000}.
These lines are useful to describe the stellar pulsation but they form at the stellar photosphere and its surroundings, where part of the expelled matter is still gravitationally bound to the star.
On the contrary, the pure rotational lines of H$_2$ are expected to be mostly excited throughout the gas acceleration zone and beyond, where the gas is not bound to the star.
\citet{keady_1993} predicted that the pure rotational line S(1) displays a P Cygni profile in \irc{} with absorption and emission components of $\simeq 2\%$ and $1\%$ the continuum, respectively.
Despite this relatively strong intensity for the high sensitivity detectors currently available, this line had remained elusive thus far.

The method commonly used to determine the mass-loss rate of an evolved star is based on the pure rotational spectrum of CO \citep[e.g.,][]{hofner_2018}.
However, the CO abundance relative to H$_2$ is poorly constrained from observations and standard values that range from a few times $10^{-4}$ to $3\times 10^{-3}$ are usually assumed \citep{groenewegen_1998,teyssier_2006,ramstedt_2008,debeck_2010,guelin_2018}.
Chemical models indicate that the order of magnitude for the CO abundance is right \citep[e.g.,][]{agundez_2020} but the uncertainties on the abundances of the chemical elements in the stellar atmosphere derived from observations can be of a factor of 2 to 3.
These are particularly important for envelopes where the central star is strongly obscured by dust or optically thick bands of abundant photospheric molecules such as C$_2$, CN, or TiO$_2$ \citep[e.g.,][]{lambert_1986}.

In this Letter, we report the first detection of the S(1) line of the pure rotational spectrum of H$_2$ toward an AGB star, \irc, based on high spectral resolution observations carried out with the Echelon-cross-echelle Spectrograph \citep[EXES;][]{richter_2018} mounted on the  Stratospheric Observatory for Infrared Astronomy \citep[SOFIA;][]{temi_2018}.

\section{Observations}
\label{sec:observations}

SOFIA/EXES observations aimed to observe the H$_2$ rotational lines S(1), S(5), and S(7), at 17.0348, 6.9095, and 5.5112~$\mu$m, respectively \citep[e.g., the HITRAN database;][]{gordon_2017}, which are available thanks to the SOFIA ability to minimize the atmospheric opacity around these wavelengths.
The S(3) line is blended with a telluric O$_3$ band and cannot be easily observed with SOFIA.

\irc{} was observed on three separate flights.
The first and second set of observations were performed on 26 and 27 Oct 2018, which corresponded to Doppler shifts of $-47.5$~\kms.  
The third set were performed on 26 Apr 2019 for a better Doppler shift for the H$_2$ S(1) setting ($+8.4$~\kms).
SOFIA was at altitude of 13.4~km for the first two flights and 11.9~km for the third flight.

The H$_2$ S(1), S(5), and S(7) lines were observed with three EXES ``High-Medium'' settings centered near 587, 1447, and 1814~\cm{} (17.04, 6.91, and 5.51~$\mu$m), respectively.
All settings were carried out on each flight and used the EXES 1\farcs9 wide slit.
The slit length for the S(1) setting was $\simeq 19\arcsec$ and we nodded \irc{} along the slit.
The slit lengths for S(5) and S(7) were about 10\arcsec and 8\arcsec, respectively, and \irc{} was nodded on and off the slit.
The resolving power ranged from 88,000 at 5.5~$\mu$m to 95,000 at 17~$\mu$m, which implies spectral resolutions around 3.3~\kms.

\begin{figure}
  \centering
  \includegraphics[width=0.475\textwidth]{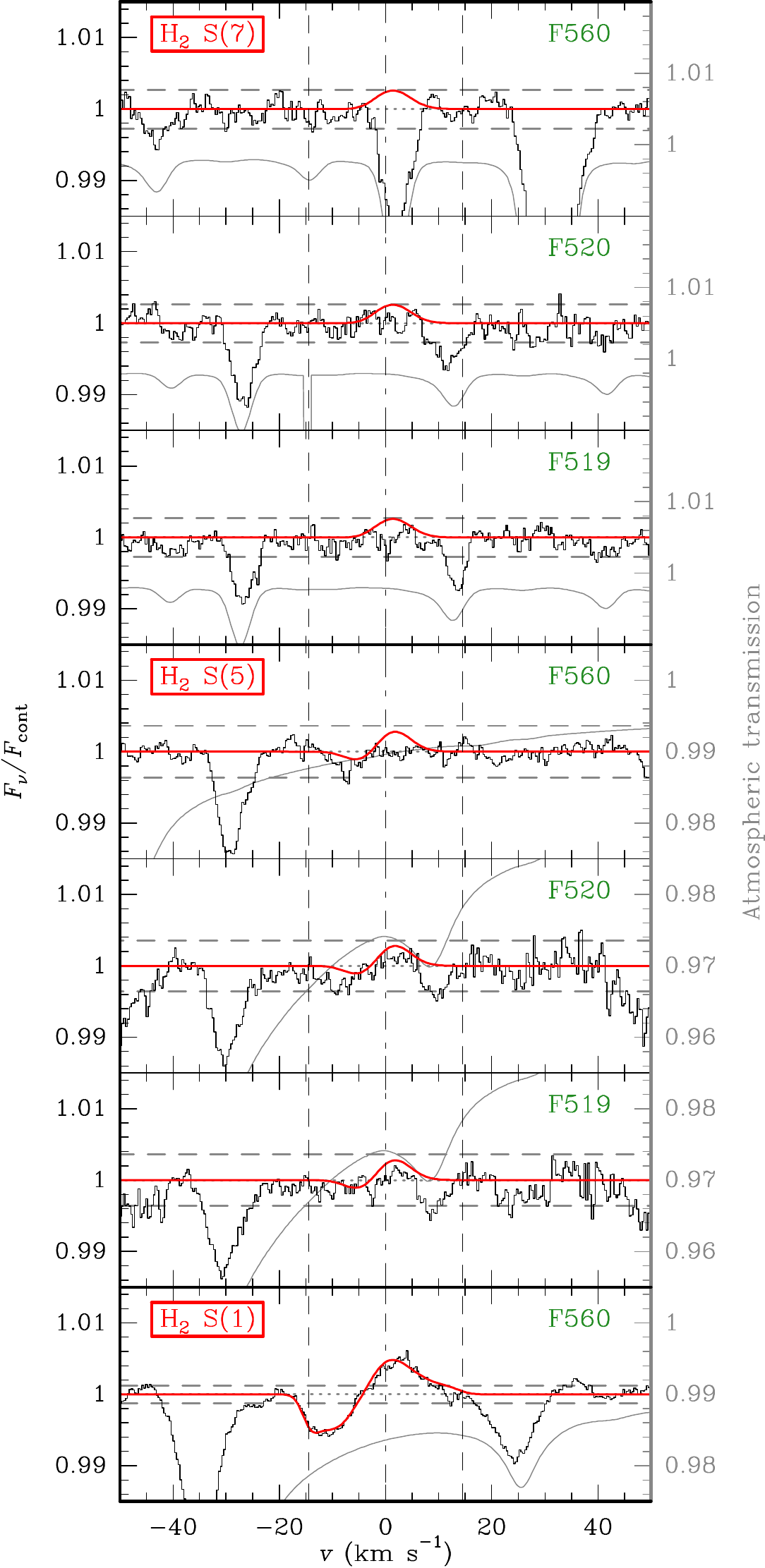}
  \caption{H$_2$ lines observed toward \irc{} (black histograms) and models (red curves).
    The model for the S(1) line is the best fit (see text).
    The gray spectra are the atmospheric transmission.
    The flights are given in green at the right top of each panel.
    F560 was the only flight where the S(1) line was not strongly blended with a close strong telluric line.
    The vertical dotted-dashed line at zero velocity represents the systemic velocity of \irc{} and the vertical gray dashed lines at $\pm 14.5$~\kms{} the terminal gas expansion velocity in front and behind the star.
    The horizontal gray dashed lines are at the $3\sigma$ level.
  }
  \label{fig:f1}
\end{figure}

\begin{figure}
  \centering
  \includegraphics[width=0.475\textwidth]{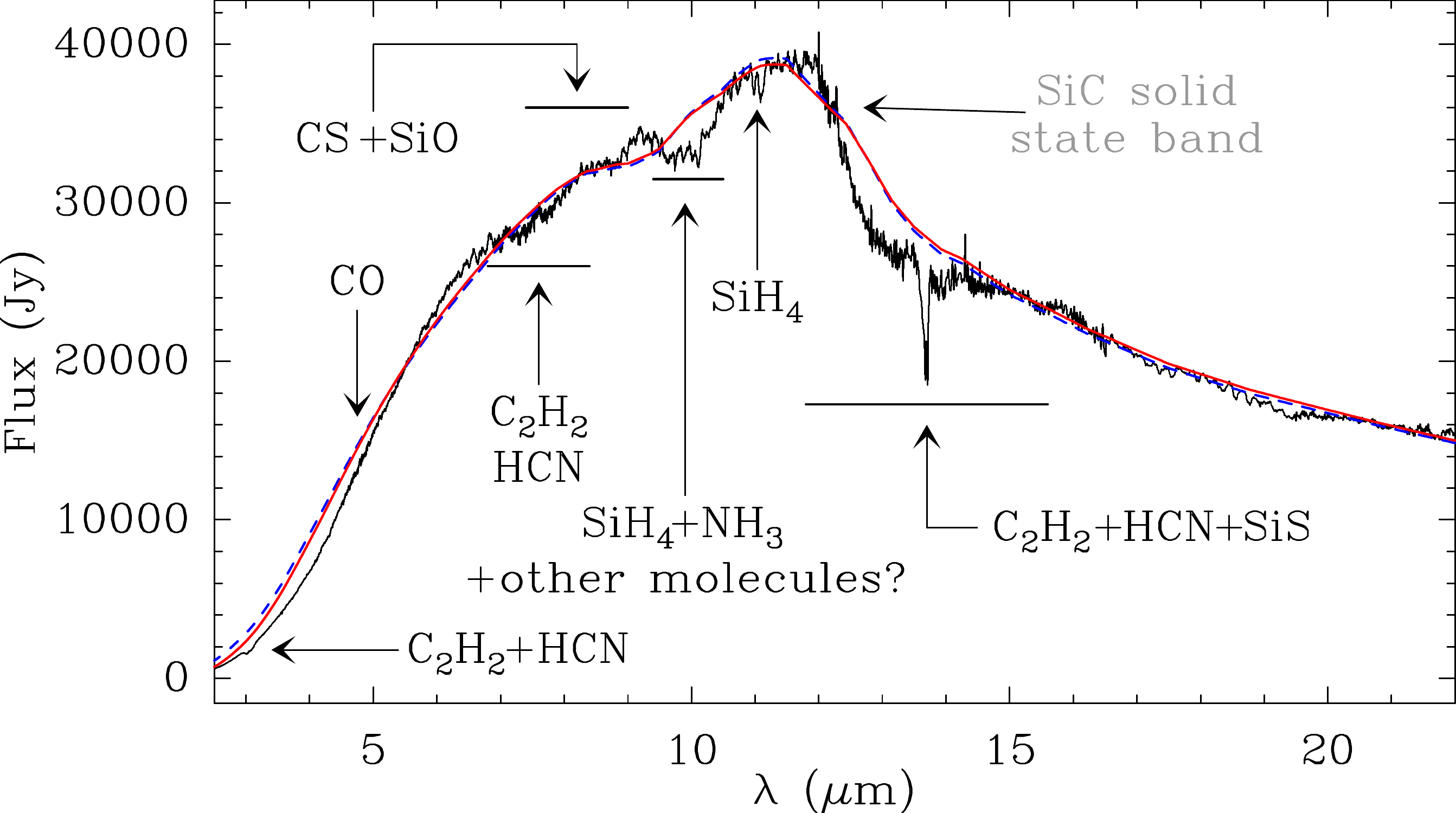}
  \caption{\textit{ISO}/SWS observed and fitted continuum emission of \irc{} (black and red).
    The continuum emission calculated assuming no dust inwards from 5\rstar{} is plotted as a dashed blue curve.
    Molecular and solid state bands are indicated \citep[e.g.,][]{fonfria_2021}.
    }
  \label{fig:f2}
\end{figure}

All EXES data were reduced using the Redux pipeline \citep{clarke_2015}. 
The baselines of the observed spectra were removed by cubic-spline fitting them aiming to keep the profile of the features coming from \irc{} intact.
The RMS noise for the settings at 587, 1447, and 1814~\cm{} are 0.04\%, 0.12\%, and 0.09\% of the continuum, respectively.
The observed spectra around the rest frequencies of the target H$_2$ lines are shown in Fig.~\ref{fig:f1}.
The adopted systemic velocity for \irc{} was $\simeq -26.5$~\kms{} \citep[e.g.,][]{cernicharo_2000}.
SOFIA/EXES observations were complemented with the low spectral resolution spectrum of \irc{} acquired with \textit{ISO}/SWS (\citealt{cernicharo_1999}; Fig.~\ref{fig:f2}).

\section{Results and discussion}
\label{sec:results}

\begin{figure*}
  \centering
  \includegraphics[width=\textwidth]{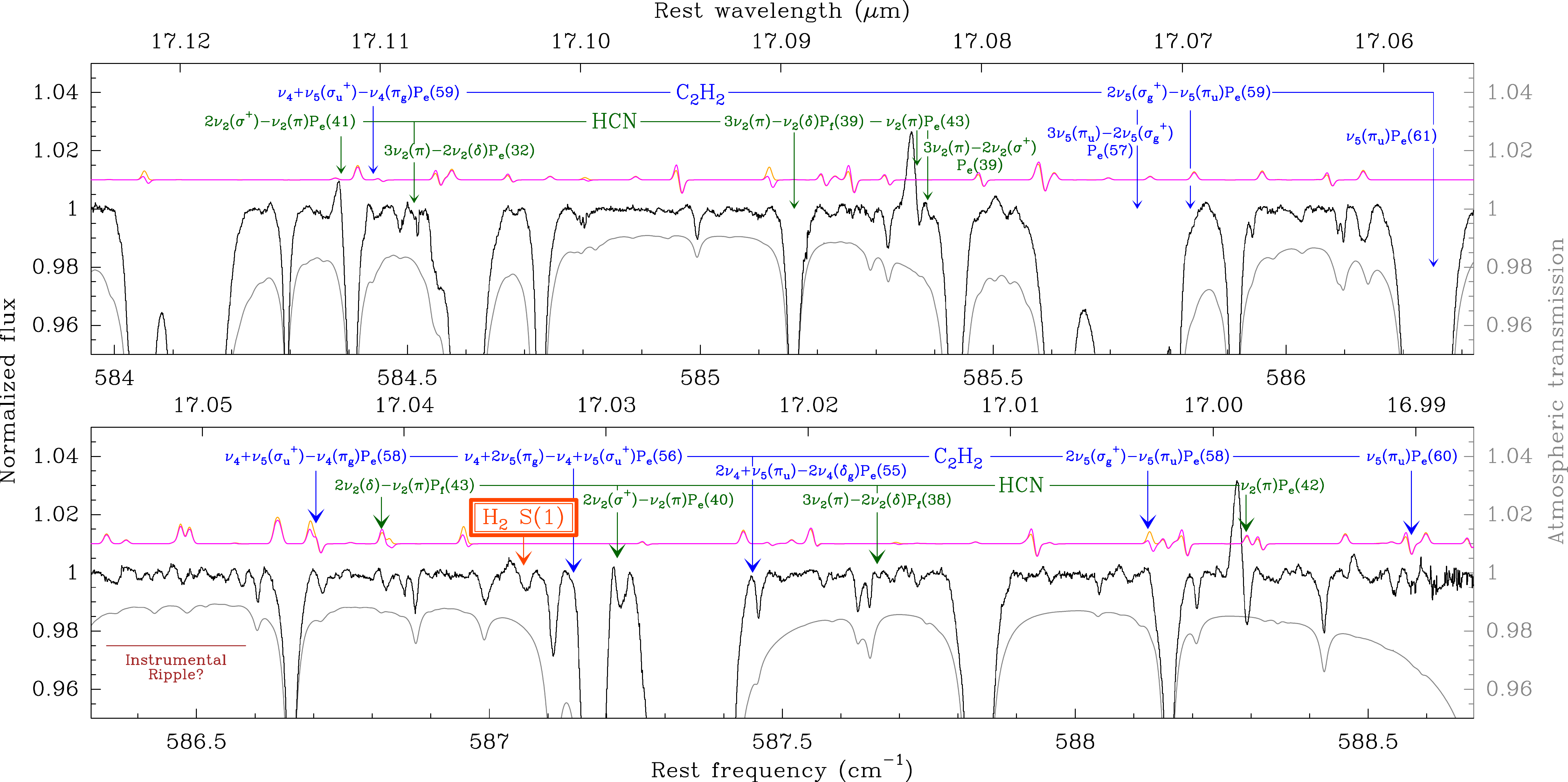}
  \caption{Observed spectrum around the rest frequency of the S(1) line of ortho-H$_2$ (labeled in red).
    The lines of C$_2$H$_2$ and HCN in this range are indicated in blue and green, respectively.
    Approximate spectra of HNC are plotted in orange and magenta (see text).
    The gray spectrum is the atmospheric transmission calculated with ATRAN \citep{lord_1992}.
}
  \label{fig:f3}
\end{figure*}

The spectral range observed to detect the S(1) line shows a low line density with several weak lines in addition to telluric features (Fig.~\ref{fig:f3}).
The strongest lines coming from \irc{} in this range are of the $\nu_2$ fundamental band of HCN.
Other weaker lines can be assigned to hot bands of HCN and C$_2$H$_2$.
The range covers the high-$J$ part of the infrared spectrum of HNC.
The abundance of this molecule in \irc{} is low \citep{cernicharo_2013} and the lines are expected to be quite weak.
However, the frequencies of several of the unidentified lines are compatible with some HNC lines and might be produced by this molecule.
The expected intensity of the HNC spectrum is analyzed in Appendix~\ref{sec:hnc.modeling}.

The H$_2$ S(1) line is located in the wing of a strong telluric line, which was corrected with a low degree polynomial.
The H$_2$ line displays a P Cygni profile where the absorption component peaks at a Doppler velocity of $-12$~\kms{} with respect to the systemic velocity and its full width at half depth is $\simeq 8.5$~\kms.
The emission component peaks at $+3$~\kms{} but this can be an effect of noise.
Its overall shape suggests that it peaks at $\simeq +1.2$~\kms, which is compatible with gas emission at the systemic velocity considering the spectral resolution.
Its full width at half maximum is $\simeq 8.2$~\kms.
The intensities of both components are $\simeq 0.5\%$ of the continuum (S/N~$\simeq 12$).
The maximum and minimum Doppler velocities, estimated by fitting the wings of the line profile with exponentials, are $12.4\pm 1.4$ and $-17.6\pm 0.5$~\kms.
The minimum Doppler velocity at half depth is $-15.5\pm 0.7$~\kms.
These velocities are compatible with ejected gas at the canonical terminal expansion velocity for \irc{} of $14.5$~\kms{} \citep{cernicharo_2000}, a line width (thermal+turbulent) of $1-3$~\kms{} for the fully accelerated H$_2$, and the spectral resolution of the observations.

The other two observed lines, S(5) and S(7), are not present above the detection limit in the corresponding settings.
The spectral ranges where they should be do not show persistent features at the expected frequencies after being observed during three flights.

\subsection{Modeling of H$_2$ lines and dust opacity}
\label{sec:line.modeling}

The H$_2$ S(1) line and the continuum emission of \irc{} were fitted with the code developed by \citet{fonfria_2008,fonfria_2014} to calculate the emission of a CSE composed of molecular gas and dust.
The use of this code permitted us to derive the mass-loss rate, the kinetic temperature of the gas, $\tk(r)$, and the gas expansion velocity profile, $\vexp(r)$, where $r$ is the distance from the star, together with the dust grains distribution and their temperature across the envelope.
More information about the fitting procedure and envelope model is included in Appendix~\ref{sec:envelope.model}.

We initially assumed that no dust exists in the region of the envelope from the photosphere to 5\rstar{} \citep[e.g.,][]{ridgway_1988,keady_1993,fonfria_2008,fonfria_2015}.
However, the synthetic S(5) and S(7) lines were stronger than S(1) in this case.
No variation of the envelope model explained the presence of S(1) and the observed absence of S(5) and S(7) simultaneously except the addition of dust in this region.
Strong dust formation seems to occur around 5\rstar{} but dust might form closer to the star as well, as suggested by \citet{danchi_1990}, \citet{menshchikov_2001}, and \citet{agundez_2020}.
For the sake of simplicity, we assumed that dust grains start to form at the stellar photosphere and their density linearly grows up to 5\rstar{} (Appendix~\ref{sec:envelope.model}).
In this scenario, a density $\gtrsim 7.0\times 10^{-3}$~\cmmm{} at the photosphere resulted in synthetic S(5) and S(7) lines below the detection limit while the S(1) line was barely affected.
This is a consequence of the much lower dust opacity at longer wavelengths and the larger extent of the region where the S(1) line forms.
The S(0) and S(2) lines of para-H$_2$ (not observed) are predicted to have peak intensities of 0.1\% and 0.2\% of the continuum, respectively, which are very close to the EXES detection limit or below it.

\subsection{Results of the fit to the S(1) line}
\label{sec:bestfit}

The lack of a permanent dipole moment due to the homonuclearity of H$_2$, implies that electric dipole transitions are not allowed.
Consequently, the rotational levels of any vibrational state of the electronic ground state are populated only by collisions throughout the whole envelope.

\begin{figure}
  \centering
  \includegraphics[width=0.475\textwidth]{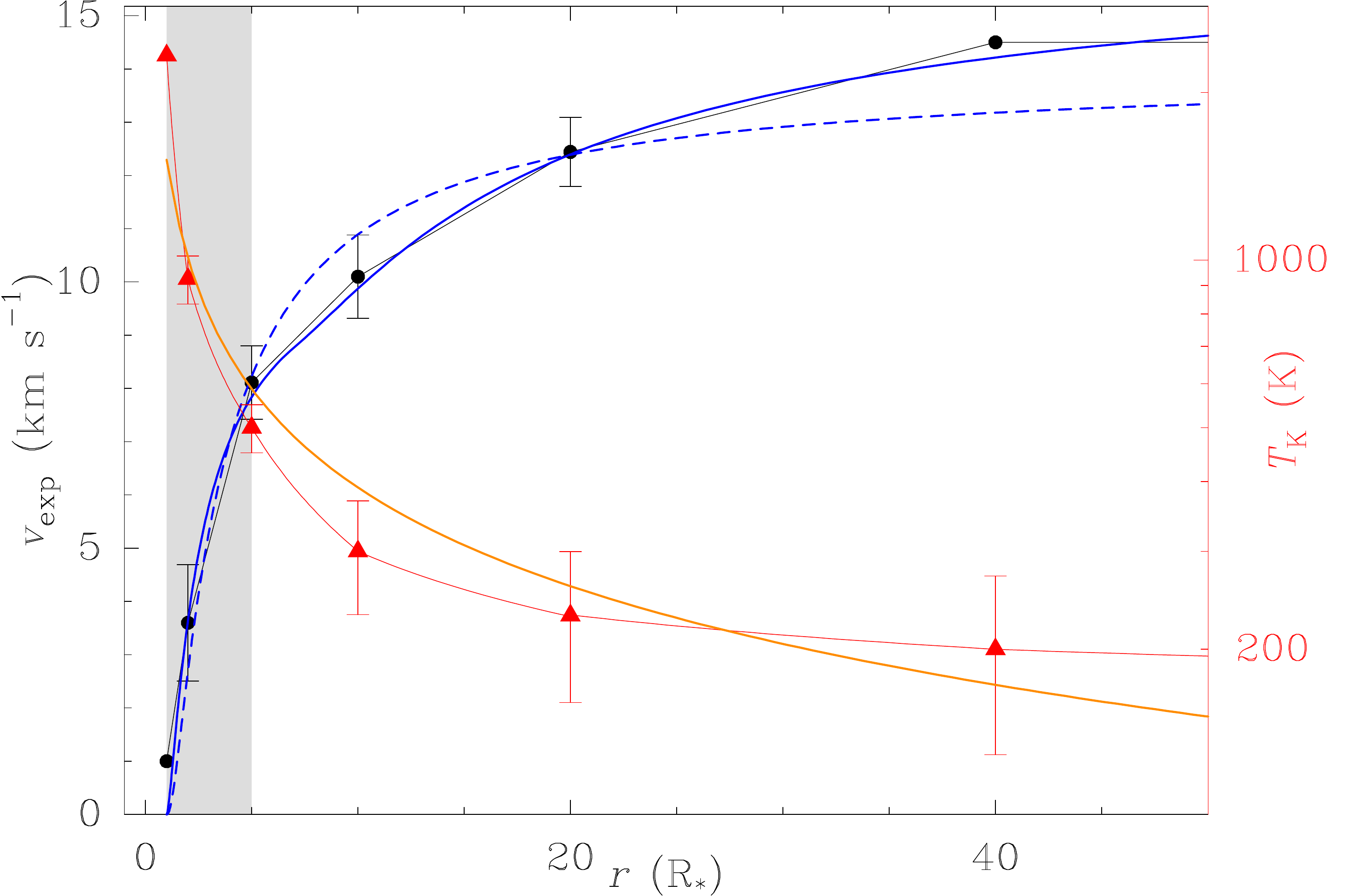}
  \caption{Gas expansion velocity and kinetic temperature derived from the fit to the H$_2$ S(1) line (black and red, respectively).
    The dashed blue curve is a fit to the gas expansion velocity assuming the theoretical profile for an expansion envelope driven by radiation pressure on dust grains ($v_\subscript{exp}=v_\infty\left(1-R_0/r\right)^\beta$, where $v_\infty=14.4$~\kms, $R_0=1\rstar$, and $\beta=2.4$).
    The solid blue curve is the same but considering two different acceleration regimes ($v_{\infty,1}=11.3$~\kms, $R_{0,1}=1\rstar$, $\beta_1=1.6$, $v_{\infty,2}=5.2$~\kms, $R_{0,2}=7\rstar$, and $\beta_2=2.2$).
    The orange profile is a fit to \tk{} assuming a power-law.
    This quantity is uncertain in the range $1-5\rstar$, shadowed in gray (see text).
  }
  \label{fig:f4}
\end{figure}

The free parameters of the model were the kinetic temperature and gas expansion velocity at fixed positions from 1 to 60\rstar, and the mass-loss rate.
The \tk{} and \vexp{} between adjacent positions were interpolated with power and linear laws, respectively, depending on $r$.
All these parameters are essentially independent and have different effects on the line profile.
The mass-loss rate was accurately determined since it mostly affects the intensity of the emission and absorption components.
The result of the fitting process to the S(1) line can be seen in Fig.~\ref{fig:f1} and the kinetic temperature and gas expansion velocity profiles are plotted in Fig.~\ref{fig:f4}.
Our envelope model suggests that the S(1) line traces H$_2$ within $\simeq 150\rstar$ from the star and only the emission component is slightly sensitive to changes in the physical conditions at the stellar photosphere.

The derived mass-loss rate is $(2.43\pm 0.21)\times 10^{-5}$~\mlr.
\citet{guelin_2018} obtained a rate of $(2.7\pm 0.5)\times 10^{-5}$~\mlr{} based on an assumed CO abundance relative to H$_2$ of $6\times 10^{-4}$.
Since both rates must be equal, the CO abundance compatible with our observations is $(6.7\pm 1.4)\times 10^{-4}$.
We note that the uncertainty is significantly improved with respect to theoretical estimates.
\citet{guelin_2018} used the interferometric CO emission at the systemic velocity, which is not affected by the position and thickness of the photodissociation region, contrary to other estimates based on single-dish observations \citep{saberi_2019}.
Hence, it is possible to linear scale the optical depths.
Nevertheless, other mass-loss rates in the range $1.0-5.0\times 10^{-5}$~\mlr{} were proposed or used in the past \citep{schoier_2001,teyssier_2006,ramstedt_2008}.
These rates were based on CO abundances relative to H$_2$ of $0.6-2.9\times 10^{-3}$ previously determined for \irc{} or samples of C-rich stars \citep{olofsson_1993,crosas_1997,groenewegen_1998}.
Our CO abundance is compatible with the lower limit of this range.

SiO is the most abundant oxygen-bearing molecule in \irc{} after CO.
Its abundance is $0.3-2.0\times 10^{-7}$ \citep{velilla-prieto_2019}, which is $\gtrsim 3000$ times lower than for CO.
The amount of oxygen in dust grains is also negligible, since refractory oxygen-bearing molecules are not expected to form in the inner layers of C-rich envelopes \citep{agundez_2020}.
Hence, the O/H abundance ratio is half the CO abundance with respect to H$_2$, $(3.3\pm 0.7)\times 10^{-4}$.
This means that [O/H]=$\log(\textnormal{O/H})-\log(\textnormal{O/H})_\odot=-0.16\pm 0.10$ if $\log(\textnormal{O/H})_\odot=-3.31\pm 0.05$ is adopted \citep{asplund_2009}.

The C/H ratio can be calculated by summing the abundances relative to H$_2$ of the most abundant carbon-bearing molecules (CO, C$_2$H$_2$, and HCN).
The abundances of C$_2$H$_2$ and HCN are $(1.5\pm 0.5)\times 10^{-4}$ and $(6\pm 3)\times 10^{-5}$, respectively \citep{fonfria_2008}, after they are corrected for the distance to the star and the mass-loss rate.
The abundance of the next most abundant molecule, CH$_4$, is $7\times 10^{-6}$ \citep[corrected;][]{keady_1993} and can be neglected.
The C/H abundance ratio for the gas phase is thus $(5.2\pm 0.9)\times 10^{-4}$.
A significant fraction of the ejected carbon likely condenses onto dust grains so this ratio is a lower limit and [C/H]$>0.28\pm 0.09$ in the stellar atmosphere \citep[$\log(\textnormal{C/H})_\odot=-3.57\pm 0.05$;][]{asplund_2009}.
Therefore, the C/O ratio is $>1.5\pm 0.4$.
The O/H and C/H ratios are compatible with those derived for other C-rich stars \citep{lambert_1986}.
The C/O ratio is in the range $1.2-2.0$, which was established for average C-rich stars \citep{winters_1994} and encompasses the values commonly used in theoretical works on circumstellar chemistry \citep[e.g.,][]{willacy_1998,agundez_2020}.
\irc{} would thus be the envelope of an evolved common C-rich star.

The gas expansion velocity profile throughout the envelope of an AGB star is expected to be described by the function $v_\subscript{exp}(r)=v_\infty\left(1-R_0/r\right)^\beta$, where $v_\infty$ is the terminal velocity, $R_0$ the radius where the acceleration starts, and $\beta$ is typically around 0.5 for expanding envelopes of cool stars \citep{decin_2006}.
With this exponent, the gas of \irc{} would reach the terminal velocity in 10\rstar.
However, an exponent of $\simeq 2.4$ is required to describe the velocity profile derived from our fit to the H$_2$ S(1) line, which means a lower gas acceleration.
In addition, the observed terminal velocity of $\simeq 14.5$~\kms{} would be reached beyond 60\rstar{} (Fig.~\ref{fig:f4}) and the gas expansion velocity from 10 to 15\rstar{} would be higher than our fit to the H$_2$ S(1) line suggests.

These discrepancies can be solved if we consider an additional acceleration regime starting at $\simeq 7\rstar$ from the star.
This improves the agreement between 10 and 15\rstar{} and the gas reaches the terminal expansion velocity before $\simeq 40\rstar$.
However, the large region of the envelope that contributes to the line absorption at the terminal velocity ($\simeq 150\rstar$) makes our estimate of this position uncertain.
Previous estimates \citep[$\simeq 20\rstar$;][]{keady_1993,fonfria_2008,fonfria_2015,agundez_2012,cernicharo_2013} are more reliable since they are based on lines that are more sensitive to the physical conditions in the 15$-$30\rstar{} region of the envelope.

The slow gas acceleration can be explained if dust grains keep growing over tens of stellar radii or the gas density is irregular due to, for instance, the presence of denser or diluted shells, which may hamper or favor the acceleration of dust grains.
Considering the complexity of \irc{} in the intermediate and outer layers \citep{cernicharo_2015,guelin_2018}, similar structures in the gas and dust distributions might exist in the inner envelope.

The \tk{} decreases quickly during the first 10\rstar{} to adopt a more moderate dependence with the distance from the star that turns into the power-law proposed by \citet{guelin_2018} for the intermediate and outer envelope.
We note that \tk{} might not be well described in the range $1-5\rstar$ due to the uncertainties related to hot dust.
The slower decrease of \tk{} beyond 10\rstar{} might indicate a lower cooling rate related to a milder dust opacity decrease over the distance from the star caused, for instance, by the dust grains growth.
A fit to the kinetic temperature with a simple power-law $\propto r^{-\alpha}$ indicates that it can be described with $\alpha\simeq 0.59$, similar to previous estimates \citep[e.g.,][]{agundez_2012,debeck_2012}.

\section{Summary and conclusions}
\label{sec:summary}

In this Letter, we present new high spectral resolution observations of the S(1), S(5), and S(7) lines of ortho-H$_2$ at 17.04, 6.91, and 5.51~$\mu$m, respectively, toward the C-rich AGB star \irc.
The S(1) line was detected in the wing of a strong telluric feature whereas no line is noticeable at the frequencies of S(5) and S(7), probably due to the opacity of hot dust close to the star, where they form.

The S(1) line was modeled with a radiation transfer code to derive the gas expansion velocity and kinetic temperature profiles.
The mass-loss rate, which is independent of chemical assumptions, was $(2.43\pm 0.21)\times 10^{-5}$~\mlr.
This rate is close to the most recent measure based on high quality interferometer observations of CO.
The difference is expected to come from the assumed theoretical CO abundance relative to H$_2$.
A CO abundance of $(6.7\pm 1.4)\times 10^{-4}$ equals both measures.
We derived O/H and C/H abundance ratios of $(3.3\pm 0.7)\times 10^{-4}$ and $>(5.2\pm 0.9)\times 10^{-4}$, respectively.
The C/O ratio is $>1.5\pm 0.4$.

This work thus represents the first time that a pure rotational line of H$_2$ is detected toward an evolved star allowing the simultaneous determination of the mass-loss rate and the CO abundance relative to H$_2$ from observations.
This detection of H$_2$ was possible thanks to the ability of SOFIA to significantly reduce the opacity of the atmosphere, a favorable Doppler shift, and the incomparable characteristics of the high spectral resolution spectrograph EXES.
Further applications of this method to other bright evolved stars open the door to improve the quality of their mass-loss rates, which can be biased due to inaccurate CO abundances, and of the CO abundances themselves.

\section*{Acknowledgments}

We thank the anonymous referee for their useful and appropriate comments.
The research leading to these results has received funding support from the European Research Council under the European Union's Seventh Framework Program (FP/2007-2013) / ERC Grant Agreement n.~610256 NANOCOSMOS.
MJR and EXES observations are supported by NASA cooperative agreement 80NSSC19K1701.
Based on observations made with the NASA/DLR Stratospheric Observatory for Infrared Astronomy (SOFIA).
SOFIA is jointly operated by the Universities Space Research Association, Inc. (USRA), under NASA contract NNA17BF53C, and the Deutsches SOFIA Institut (DSI) under DLR contract 50 OK 0901 to the University of Stuttgart.

\facility{\textit{Facility:} SOFIA(EXES)}
\newline\indent\textit{Software:} Redux \citep{clarke_2015}

\appendix

\section{Modeling of the HNC spectrum}
\label{sec:hnc.modeling}

The HNC spectrum was modeled with the radiation transfer code presented in Section~\ref{sec:line.modeling}.
We adopted the envelope model derived from the fit to the H$_2$ S(1) line assuming two different rotational temperature profiles ($T_{\subscript{rot},1}=2300(1\rstar/r)^{1.05}$ and $T_{\subscript{rot},2}=2300(1\rstar/r)^{0.6}$~K) and a vibrational temperature of  $\tvib=1200(1\rstar/r)^{0.8}$~K.
These profiles are based on the kinetic temperature derived in the current work and the rotational and vibrational temperatures for HCN of \citet{fonfria_2008}.
A 1\rstar-thick HNC layer located beyond the photosphere was adopted.
Dust was assumed to exist between the photosphere and 5\rstar{} (Section~\ref{sec:line.modeling}, Appendix~\ref{sec:envelope.model}).
The spectra obtained with an HNC abundance of $1\times 10^{-7}$ show lines with intensities similar to or slightly higher than the unidentified features noticeable in our observations (Fig.~\ref{fig:f3}).
The differences that exist between the spectra produced with the two rotational temperature profiles ($T_{\subscript{rot},1}$ in orange and $T_{\subscript{rot},2}$ in magenta in Fig.~\ref{fig:f3}) illustrate their mild influence on the shape of the HNC lines formed close to the stellar atmosphere, which is a consequence of the narrowness of the adopted HNC shell.
A high number of lines appear in emission in the synthetic spectra with no observational counterpart, which indicates that the adopted vibrational temperature could be too high for the vibrational excited states if some of the observed features can be eventually assigned to HNC.
Since the calculated HNC lines could only (tentatively) explain a fraction of the unidentified features of the spectrum, these features could be actually produced by other molecules and the HNC abundance could be even lower.
Moreover, the uncertainties involved in the existence of hot dust close to the star affects our abundance determination.
Therefore, we estimate an upper limit for the HNC abundance close to the star of $0.5-1.0\times 10^{-7}$, which is less than one order of magnitude lower than the peak HNC abundance proposed by \citet{cernicharo_2013} and \citet{agundez_2020}.

\section{Envelope model and fitting procedure}
\label{sec:envelope.model}

In our models, several quantities were adopted as fixed: a distance of 120~pc \citep{groenewegen_2012}, a stellar radius of 0\farcs02 \citep{ridgway_1988}, a solar He abundance with respect to H$_2$ of 0.17 \citep{asplund_2009}, and the local line width, which was adopted to be the combination of the thermal line width for H$_2$ and a contribution related to gas turbulence.
This latter contribution was chosen to be equal to 5~\kms{} at the stellar photosphere, decreased linearly down to 1~\kms{} at 5\rstar, and remained constant beyond \citep{agundez_2012}.
We assumed \tk{} at the photosphere equal to the effective stellar temperature \citep[$\simeq 2300$~K;][]{ridgway_1988} and to $160.3(80\rstar/r)^{0.68}$~K beyond 80\rstar{} \citep[][]{guelin_2018}, which was determined for the region of the envelope between 40 and 800\rstar{} from interferometer CO observations at the systemic velocity.
We used the shell that extends from 40 to 80\rstar{} to guarantee a smooth coupling between the dependences of the \tk{} close and far from the star.
We note that single-dish observations of the H$_2$ S(1) line are not appropriate to describe \tk{} in the intermediate and outer envelope as only a small fraction of the line is affected by the physical conditions beyond 40\rstar.
These quantities are well established and their use improved the accuracy of the parameters derived from the fit to the H$_2$ line.

We initially adopted the dust distribution over the envelope derived by \citet{fonfria_2008,fonfria_2015}, i.e., $n_\subscript{dust}(r< 5\rstar)$=0~\cmmm, $n_\subscript{dust}(r\ge 5\rstar)$=$6.1\times 10^{-3}(5\rstar/r)^2$~\cmmm, and $T_\subscript{dust}(r\ge 5\rstar)$=$825(5\rstar/r)^{0.39}$~K, considering small grains with a size of 0.05~$\mu$m and composed of 92.5\% amorphous carbon and 7.5\% amorphous SiC.
This model reproduced reasonably well the continuum observed with \textit{ISO} even at short wavelengths despite these authors did not consider dust between the star and 5\rstar{} (Fig.~\ref{fig:f2}), which was supported by previous works \citep[e.g.,][]{ridgway_1988,keady_1993}.
However, it is not clear that dust does not form closer to the star, as it is pointed out by, for instance, \citet{danchi_1990} and \citet{menshchikov_2001}.
Hence, in the current work, we explored the possibility of filling this gap with hot dust to assess its effect on the H$_2$ lines.
The density of dust grains, $n_\subscript{dust}(r)$, was assumed to linearly grow from the photosphere until 5\rstar{} and the temperature of the dust grains, $\td(r)$, was initially extrapolated from the profile beyond 5\rstar.
The value of the density at the photosphere was increased and the rest of the parameters refined until the H$_2$ lines S(5) and S(7) were below the detection limit (Fig.~\ref{fig:f1}).
The dust grain density resulted to be $7.0\times 10^{-3}$~\cmmm{} at the stellar photosphere and to follow the power-law $6.0\times 10^{-3}(5\rstar/r)^2$~\cmmm{} beyond 5\rstar.
The \td{} was $\simeq 1550$~K at the photosphere, decreasing as $\propto r^{-0.4}$.
The \textit{ISO} observations are mostly insensitive to the existence of dust in the $1-5\rstar$ region of the envelope since the subtended solid angle is too small compared to the half power beam width of \textit{ISO}.
Nevertheless, H$_2$ lines might help us constrain the density of dust grains in this region of the CSE owing to the higher inherent spatial resolution of the SOFIA/EXES observations.

Once the distribution of dust grains is determined, the S(1) line was automatically fitted several times by minimizing the $\chi^2$ function with the Gradient Descent method starting from different initial sets of parameters.
These sets were randomly calculated from the approximate parameters derived from a good fit evaluated by eye.
The minimum average deviation of the automatic fits with respect to the observed line is $\simeq 0.06\%$ of the continuum, that is $1.5\sigma$, where $\sigma$ is the RMS noise ($\simeq 0.04\%$).
Further information about the latest version of the code, the fitting methodology, and the uncertainties calculation can be found in \citet{fonfria_2015,fonfria_2021}.

\end{document}